\documentstyle[12pt,epsf]{article}
\newcommand{\nn}{\nonumber}
\newcommand{\be}{\begin{equation}}
\newcommand{\ee}{\end{equation}}
\newcommand{\bea}{\begin{eqnarray}}
\newcommand{\eea}{\end{eqnarray}}
\def\bfnabla{\mbox{\boldmath $\nabla$}}

\def\bfsigma{\mbox{\boldmath $\sigma$}}
\def\lQ{\Lambda_{\rm QCD}}

\def\als{\alpha_{\rm s}}
\def\siml{{\ \lower-1.2pt\vbox{\hbox{\rlap{$<$}\lower6pt\vbox{\hbox{$\sim$}}}}\ }} 
\def\vac{\hbox{vac}}
\newcommand{\Appendix}[1]%
    {%
     \section{#1}%
      }

\begin{document}
\begin{titlepage}
\begin{flushright}
\tt{CERN-TH/2000-053\\ UB-ECM-PF 00/03\\ UWThPh-1999-47}
\end{flushright}
\vspace{1cm}
\begin{center}
\begin{Large}
{\bf The QCD Potential at $O(1/m)$\\[2cm]}
\end{Large}
{\large Nora Brambilla}\footnote{nora.brambilla@cern.ch}\\
{\it Institut f\"ur Theoretische Physik, Boltzmanngasse 5, A-1090 Vienna,
  Austria and\\
\it Theory Division CERN, 1211 Geneva 23, Switzerland}
\\[0.6cm] 
{\large Antonio Pineda}\footnote{antonio.pineda@cern.ch}\\
{\it Theory Division CERN, 1211 Geneva 23, Switzerland}\\[0.6cm]
{\large Joan Soto}\footnote{soto@ecm.ub.es}\\
{\it Dept. d'Estructura i Constituents de la Mat\`eria and IFAE, U. Barcelona \\ 
     Diagonal 647, E-08028 Barcelona, Catalonia, Spain}\\[0.6cm]
{\large Antonio Vairo}\footnote{antonio.vairo@cern.ch}\\
{\it Theory Division CERN, 1211 Geneva 23, Switzerland}
\end{center}

\vspace{1cm}
\begin{abstract}
\noindent 
Within an effective field theory framework, we obtain an expression for the next-to-leading 
term in the $1/m$ expansion of the singlet $Q{\bar Q}$ QCD potential in terms of
Wilson loops, which holds beyond perturbation theory.  The ambiguities in 
the definition of 
the QCD  potential beyond leading order in $1/m$ are discussed and a specific
 expression for the $1/m$ potential is given.
We explicitly evaluate
this expression at one loop and compare the outcome with the existing
perturbative results. 
On general grounds we show that for quenched QED and fully Abelian-like
models this expression exactly vanishes.
\vspace{5mm} \\
{\small PACS numbers: 12.39.Hg, 12.38.Lg, 12.38.Bx, 12.38.Gc, 12.39.Pn}
\end{abstract}
\end{titlepage}
\vfill
\setcounter{footnote}{0} 
\pagenumbering{arabic}

\section{Introduction}
After the discovery of the first heavy-quark bound states, the $\psi$ and the
$\Upsilon$ systems, it was soon realized that a non-relativistic picture seemed to
hold for them. This is characterized by, at least, three scales: hard (the mass $m$,
of the heavy quarks), soft (the relative momentum of the heavy-quark--antiquark 
$|{\bf p}| \sim mv$, $ v \ll 1$), and ultrasoft (US, the typical binding
energy $E \sim mv^2$ of the bound state system). It was also seen that, if one wanted to describe 
the whole spectrum of the $\psi$ and the $\Upsilon$ systems, a perturbative evaluation of the potential
was not sufficient. This triggered the investigation of these systems by all sorts of potential 
models (see \cite{rev} for some reviews), which are, in general, quite successful phenomenologically. 
Since then, a lot of effort has been devoted to obtaining the relevant potentials to be used 
in the Schr\"odinger equation of such models from QCD 
\cite{wilson,Brown,spin1,spin2,cor,cor1,latpot,thesis} 
by relating these potentials with some Wilson loops that could eventually be computed 
on the lattice or by using some vacuum models. The expression for the leading spin-independent 
potential, of $O(1/m^0)$, has been known since long and corresponds to the static Wilson loop
\cite{wilson,Brown}. The expressions for the leading spin-dependent potentials 
in the $1/m$ expansion, of $O(1/m^2)$, have been calculated in Refs. \cite{spin1,spin2}. 
The $1/m$ corrections to these potentials have proved to be very difficult to obtain.
To our knowledge the spin-independent case has been addressed only 
in Refs. \cite{cor,thesis}. The result of Ref. \cite{cor} does not 
reproduce the one-loop perturbative QCD potential (see discussion at the end of section 5.1) 
and, therefore, appears to be incomplete. In Ref. \cite{thesis} the author does not 
succeed in obtaining suitable finite expressions. We conclude, therefore, 
that the question of the $1/m$ corrections to the QCD potential has not been settled yet and 
hence deserves further studies. In this work we will present 
an {\it ab initio} and systematic calculation of the QCD potential up to $O(1/m)$. 
We will get a new expression of the $1/m$ potential that is finite, consistent 
with one-loop perturbative QCD and suitable to be evaluated by lattice simulations.  

We will perform the calculation by integrating out in two steps the hard and the soft scales  
characterizing the heavy-quark--antiquark system. This is implemented by introducing suitable effective 
field theories. This approach allows us to express the heavy-quark--antiquark dynamics in terms 
of systematic and controlled expansions. It has proved to be a powerful computational 
tool in several different situations. For instance, the hard log corrections ($\sim \ln m$) 
to the Eichten--Feinberg--Gromes potentials were computed in this way  in Ref. \cite{chen} 
(see also \cite{latpot}). Moreover, we believe that the effective field theories provide a suitable 
framework where eventually some long-standing conceptual questions will be clarified. 
In particular, the extent of validity of the na\"\i ve potential picture for the heavy 
quarkonium dynamics, assumed in potential models, could be affected by the consideration 
of extra degrees of freedom such as hybrids and pions.

The two QCD effective field theories that arise from integrating out the scales $m$ and $mv$ 
are called NRQCD and pNRQCD respectively. Non-relativistic QCD (NRQCD) was first 
introduced in Ref. \cite{Lepage}. It has an ultraviolet cut-off much smaller than the mass $m$ 
and much larger than any other scale (in particular much larger than $\lQ$, which means 
that the matching from QCD to NRQCD can always be done perturbatively \cite{Manohar,Match}). 
NRQCD has proved to be extremely successful in studying ${\bar Q} Q$ systems near threshold. 
The Lagrangian of NRQCD is organized in powers of $1/m$, making in this way explicit the 
non-relativistic nature of the described systems. The maximum size of each term can be 
estimated by assigning the soft scale to any dimensionful object. 
In order to connect NRQCD with a potential picture the degrees of freedom
with energies much larger than $mv^2$
have to be integrated out. Once this is done, one is left with 
a new QCD effective field theory called potential NRQCD (pNRQCD)
\cite{pNRQCD,long}\footnote{For related work on these issues within the perturbative regime of NRQCD 
we refer to \cite{Manohar,Match,others,pos}.}. 
Strictly speaking, pNRQCD has two ultraviolet cut-offs, $\Lambda_1$ and $\Lambda_2$. 
The former fulfils the relation  $ mv^2$  $\ll \Lambda_1 \ll$  $mv$ and is the cut-off of the
energy of the quarks, and of the energy and the momentum of the gluons, 
whereas the latter fulfils $mv \ll \Lambda_2 \ll m$ and is the cut-off of the relative
momentum of the quark--antiquark system, ${\bf p}$.  
This theory has been thoroughly studied, and its matching with NRQCD performed, in the situation
$mv \gg \lQ$. In this case the matching can be performed perturbatively. In this paper 
we will allow $\lQ$ to be as large as $mv$ and, therefore, we cannot rely on perturbation theory. 
Nevertheless, we will assume that the matching between NRQCD and pNRQCD can
be performed order by order in the $1/m$ expansion.  
We will present, for the general situation $\lQ \siml mv$,  
the matching of NRQCD to pNRQCD at the next-to-leading order in the $1/m$ 
expansion in the singlet sector (to be defined later). This will prove to be
equivalent to computing the $1/m$ potential. Formulas that are similar to
those in the classical papers of Refs. \cite{spin1,spin2,cor} will be
worked out. No further degrees of freedom with US energy besides the singlet
will be considered. This means that non-potential effects will be neglected in
this paper (in the perturbative situation this would be equivalent  
to working at zero order in the multipole expansion). 
A detailed study of the matching between NRQCD and pNRQCD in the situation
$\lQ \siml mv$, including US effects, will be worked out elsewhere. 

The paper is organized in the following way. In section \ref{quanto} we introduce NRQCD 
up to order $1/m$ and discuss its static limit.
Moreover, we define what will be pNRQCD in the present context. In sections \ref{secwilson} 
and \ref{qmmatch} we derive the $1/m$ corrections to the potential, by matching NRQCD to pNRQCD. 
The Green functions are worked out in the Wilson loop language. The $1/m$ potential can 
be written in a simple way as insertions of chromoelectric fields on a static Wilson loop. In section \ref{application} we compute the potential perturbatively up to one loop.
Quenched QED and Gaussian models of the QCD long-range dynamics are also
discussed. Finally, section 6 is devoted to the conclusions and the Appendix
to show how unitary transformations affect the form of
the potential.

\section{NRQCD and pNRQCD}
\label{quanto}
The Lagrangian of NRQCD up to order $1/m$ reads
\bea
{\cal L}_{\rm NRQCD} &=& \psi^\dagger \left( iD_0 +  {{\bf D}^2\over 2 m_1} 
+g c_F^{(1)} {\bfsigma \cdot {\bf B}\over 2 m_1}\right)\psi
+  \chi^\dagger\left(iD_0 -  {{\bf D}^2\over 2 m_2}
-g c_F^{(2)} {\bfsigma \cdot {\bf B}\over 2 m_2} \right)\chi 
\nn\\
& & - {1\over 4} G_{\mu \nu}^a G^{\mu \nu \, a} \,,
\label{nrqcdl}
\eea
where $\psi$ is the Pauli spinor field that annihilates the fermion, $\chi$
is the Pauli spinor field that creates the antifermion, $i D^0=i\partial_0-gA^0$,  
and $i{\bf D}=i\bfnabla+g{\bf A}$. The matching coefficients $c_F^{(j)} \simeq 
1 + O(\als)$ are not going to  be relevant here. For simplicity, light fermions are not
explicitly displayed. Their inclusion does not change the results of this paper,
although it changes the expression of some intermediate formulas.

The Hamiltonian associated to the Lagrangian (\ref{nrqcdl}) is 
\begin{eqnarray*}
H &=& H^{(0)}+{1 \over m_1}H^{(1,0)}+{1 \over m_2}H^{(0,1)} + \dots,
\label{HH}\\
H^{(0)} &=& \int d^3{\bf x} {1\over 2}\left( {\bf E}^a{\bf E}^a 
+{\bf B}^a{\bf B}^a \right),
\label{H0}\\
H^{(1,0)} &=& - {1 \over 2} \int d^3{\bf x} \psi^\dagger \left( {\bf D}^2 
+ g c_F^{(1)} \bfsigma \cdot {\bf B}\right) \psi,  
\qquad 
H^{(0,1)} = {1 \over 2} \int d^3{\bf x} \chi^\dagger \left({\bf D}^2 
+ g c_F^{(2)} \bfsigma \cdot {\bf B} \right) \chi,
\label{H1}
\end{eqnarray*}
and the physical states are constrained to satisfy the Gauss law. 
We are interested in the one-quark--one-antiquark sector of the Fock space. 

It will prove convenient to study the static limit. The one-quark--one-antiquark
sector of the Fock space can be spanned by
\begin{equation}
\vert \underbar{n}; {\bf x}_1 ,{\bf x}_2  \rangle^{(0)} :=  \psi^{\dagger}({\bf x}_1) \chi ({\bf x}_2)
|n;{\bf x}_1 ,{\bf x}_2\rangle^{(0)},\qquad \forall {\bf x}_1,{\bf x}_2
\label{basis0}
\end{equation}
where $|\underbar{n}; {\bf x}_1 ,{\bf x}_2\rangle^{(0)} $ is a gauge-invariant
eigenstate (up to a phase), as a consequence of the Gauss law, of $ H^{(0)}$
with energy $E_{n}^{(0)}({\bf x}_1 ,{\bf x}_2)$; 
$|n;{\bf x}_1 ,{\bf x}_2\rangle^{(0)}$ encodes the gluonic content of the state,
namely it is annihilated by $\chi^{\dagger}({\bf x})$ and $\psi ({\bf x})$ 
($\forall {\bf x}$). It transforms as a $3_{{\bf x}_1}\otimes 3_{{\bf x}_2}^{\ast}$ 
under colour $SU(3)$. The normalizations are taken as follows
\be
^{(0)}\langle m;{\bf x}_1 ,{\bf x}_2|n;{\bf x}_1 ,{\bf x}_2\rangle^{(0)} =\delta_{nm}
\ee
\be
^{(0)}\langle \underbar{m}; {\bf x}_1 ,{\bf x}_2|\underbar{n}; {\bf y}_1 ,{\bf y}_2\rangle^{(0)} =\delta_{nm}
\delta^{(3)} ({\bf x}_1 -{\bf y}_1)\delta^{(3)} ({\bf x}_2 -{\bf y}_2)\,.
\ee
Notice that since $ H^{(0)}$ does not contain any fermion field, $|n;{\bf x}_1,{\bf x}_2\rangle^{(0)}$ 
itself is also an eigenstate of $H^{(0)}$ with energy $E_{n}^{(0)}({\bf x}_1 ,{\bf x}_2)$. 
We have made it explicit that the positions  
${\bf x}_1$ and ${\bf x}_2$ of the quark and antiquark respectively are
good quantum numbers for the static solution $|\underbar{n};{\bf x}_1 ,{\bf x}_2 \rangle^{(0)}$, whereas
$n$ generically denotes the remaining quantum numbers, which are classified by the irreducible 
representations of the symmetry group $D_{\infty h}$ (substituting the parity generator by CP). 
The ground-state energy $E_0^{(0)}({\bf x}_1,{\bf x}_2)$ is usually associated to the static 
potential (see \cite{long} for important qualifications on this association) and the remaining 
energies $E_n^{(0)}({\bf x}_1,{\bf x}_2)$, $n\not=0$, are usually called gluonic excitations 
between static quarks. They can be computed on the lattice (see for instance \cite{michael}). 
Translational invariance implies that  $E_n^{(0)}({\bf x}_1,{\bf x}_2) = 
E_n^{(0)}(r)$, where ${\bf r}={\bf x}_1-{\bf x}_2$. 

The gap between different states at fixed ${\bf r}$ will depend on the dimensionless parameter 
$\lQ r$. In a general situation, there will be a set of states $\{n_{\rm us}\}$ such that 
$E_{n_{\rm us}}^{(0)}(r) \sim mv^2$ for the typical $r$ of the actual physical
system. 
We denote these states as ultrasoft. The aim of pNRQCD is to describe the behaviour of the ultrasoft 
states (pNRQCD has been introduced in \cite{pNRQCD} and discussed in different kinematic 
situations in \cite{long}). Therefore, all the physical degrees of freedom with energies larger
than $mv^2$ will be integrated out from NRQCD in order to obtain pNRQCD. 
In the perturbative situation $\lQ r \ll 1$, which has 
been studied in detail in \cite{long}, $\{n_{\rm us}\}$ corresponds to a
heavy-quark--antiquark state, in either a singlet or an octet configuration, plus gluons
and light fermions, all of them with energies of $O(mv^2)$. In a non-perturbative situation, which we will generically denote by
$\lQ r \sim 1$, it is not so clear what $\{n_{\rm us}\}$ is. One can think 
of different possibilities. Each of them will give, in principle, different
predictions and, therefore, it should be possible to experimentally discriminate
between them. In particular, one could consider the situation where, because
of a mass gap in QCD, the energy splitting between the ground state and the
first gluonic 
excitation is larger than $mv^2$, and, because of chiral symmetry breaking of QCD, 
Goldstone bosons (pions/kaons) appear. Hence, in this situation,  $\{n_{\rm us}\}$
would be the ultrasoft excitations about the static ground state (i.e. the
solutions of the corresponding Schr\"odinger equation),
which will be named the singlet, plus the
Goldstone bosons. If one switches off the light fermions (pure gluodynamics), 
only the singlet survives and pNRQCD becomes totally equivalent to a quantum-mechanical 
Hamiltonian, thus providing us with a qualitative explanation of how potential models
emerge from QCD. 
In addition, we shall see below how quantitative formulae can be provided in 
order to calculate the potentials in QCD.

In this paper, we will only study the pure singlet sector with no
reference to further ultrasoft degrees of freedom. 
In this situation, pNRQCD only describes the ultrasoft excitations about the
static ground state of NRQCD. In terms of static NRQCD
eigenstates, this means that only  $|\underbar{0}; {\bf x}_1 ,{\bf x}_2\rangle^{(0)}$ 
is kept as an explicit degree of freedom whereas $|\underbar{n}; {\bf x}_1 ,{\bf x}_2\rangle^{(0)}$ with  
$n\not=0$ are integrated out\footnote{In fact, we are only integrating out states with energies 
larger than $mv^2$ and all the states with $n\not=0$ will be understood in this way throughout the paper.}. 
This provides the only dynamical degree of freedom of the theory. It is described by means 
of a bilinear colour singlet field, $S({\bf x}_1,{\bf x}_2,t)$, which has the same quantum numbers 
and transformation properties under symmetries as the static ground state of NRQCD in the 
one-quark--one-antiquark sector. In the above situation, the Lagrangian of pNRQCD reads
\be
{\cal L}_{\rm pNRQCD} = S^\dagger 
\bigg( i\partial_0 -h_s({\bf x}_1,{\bf x}_2, {\bf p}_1, {\bf p}_2)\bigg) S, 
\label{pnrqcdl}
\ee
where $h_s$ is the Hamiltonian of the singlet (in fact, $h_s$ is only a
function of {\bf r}, ${\bf p}_1$, ${\bf p}_2$, which is analytic in the two
last operators but typically contains non-analyticities in {\bf r}), ${\bf
  p}_1= -i \bfnabla_{{\bf x}_1}$ and ${\bf p}_2= -i \bfnabla_{{\bf x}_2}$. It
has the following expansion up to order $1/m$ 
\be
h_s ({\bf x}_1,{\bf x}_2, {\bf p}_1, {\bf p}_2) = 
{{\bf p}^2_1\over 2 m_1} +{{\bf p}^2_2\over 2 m_2} + V_0(r) 
+ \left({1\over m_1} +{1\over m_2}\right) V_1(r).
\label{hss}
\ee

In this work we will present the matching between NRQCD and pNRQCD within an
expansion in $1/m$ using the static limit solution as a starting point.
Whereas this can be justified within a perturbative framework, in the
non-perturbative case the validity of the $1/m$ expansion cannot be generally
guaranteed\footnote{See \cite{DS} for an example where certain degrees of 
freedom cannot be integrated out in the $1/m$ expansion.}. We note that this
assumption is implicit in all the attempts at deriving the non-perturbative
potentials from QCD we are aware of.
Furthermore, we would like to 
emphasize the following important point.
The matching calculation can be
 done independently of what the specific counting
 in pNRQCD is. As argued in   \cite{Match,pos,long}, when
doing   a matching calculation we are 
integrating  out high energy degrees of
 freedom. Hence, the Wilson coefficients (potentials) that this
 integration produces are independent of
what the low energy dynamics is. The
1/m expansion just provides a convenient way to organize the matching
  calculation. One should not
 conclude that the relative size of the
 different potentials in the pNRQCD
 Lagrangian is trivially dictated by the 1/m  expansion.

\section{The Wilson loop matching}
\label{secwilson}
In this section we will work out the matching up to order $1/m$,  
in a language that is close to the traditional approach of Eichten and
Feinberg \cite{spin1,thesis}. The matching between NRQCD and pNRQCD is done by 
enforcing suitable Green functions  to be equal in the two  theories at the desired order 
in $1/m$. In this section we shall consider space-time Green functions, 
which are more conventional in non-perturbative studies.

Let us consider an interpolating state with a non-vanishing overlap with the 
ground state,   
\be
\psi^\dagger({\bf x}_1)\phi({\bf x}_1,{\bf x}_2) \chi({\bf x}_2) \vert \vac \rangle, 
\label{interfield}
\ee
where $\phi$ may in principle be everything that makes the above state overlap 
with the ground state $| \underbar{0} \rangle^{(0)}$. 
We will use here the following popular choice\footnote{With this choice we assume that 
the ground state has the $\Sigma_g^+$ quantum numbers (see \cite{long}). 
This is so in perturbative QCD as well as non-perturbative QCD according to the
available lattice simulations.}:
\begin{equation}
\phi({\bf y},{\bf x};t) \equiv {\rm P} \exp \left\{ ig \displaystyle 
\int_0^1 \!\! ds \, ({\bf y} - {\bf x}) \cdot {\bf A}({\bf x} - s({\bf x} - {\bf y}),t) \right\},
\label{schwinger}
\end{equation}
where  P is a path-ordering operator. We also define 
$\phi({\bf y},{\bf x};t=0) \equiv \phi({\bf y},{\bf x})$. 
It is going to be useful to introduce $a_n({\bf x}_1,{\bf x}_2)$, defined by 
$$
\psi^\dagger({\bf x}_1)\phi({\bf x}_1,{\bf x}_2) \chi({\bf x}_2) \vert \vac \rangle = 
\sum_n a_n({\bf x}_1,{\bf x}_2)\vert \underbar{n};{\bf x}_1,{\bf x}_2 \rangle^{(0)}. 
$$
The states $\vert \underbar{n} \rangle^{(0)}$ being defined up to a phase, 
it is convenient to fix them in such a way that all the coefficients $a_n$ are real. 
The identification of the singlet from the state (\ref{interfield})  
depends on the interpolating field used in NRQCD. This dependence is 
reflected in different normalization factors $Z$. The matching condition reads
\begin{equation}
\chi^\dagger({\bf x}_2,t) \phi({\bf x}_2,{\bf x}_1;t) \psi({\bf x}_1,t) =
Z^{1/2}({\bf x}_1,{\bf x}_2,{\bf p}_1, {\bf p}_2)  S({\bf x}_1,{\bf x}_2,t).  
\label{Sdef}
\end{equation}
As in $h_s$, the normalization factor $Z$ only depends on ${\bf r}$, ${\bf p}_1$
and ${\bf p}_2$, and is given in the form of an expansion in $1/m$ as 
\begin{equation}
Z({\bf x}_1,{\bf x}_2,{\bf p}_1, {\bf p}_2)
= Z_0(r) + \left({1\over m_1}+ {1\over m_2}\right) Z_1(r)
+ i Z_{1,p}(r) {\bf r}\cdot \bigg( {{\bf p}_1 \over m_1}-{{\bf p}_2 \over m_2}
\bigg) + \dots, 
\label{zed}
\end{equation}
where we have made use of the fact that the NRQCD Lagrangian (\ref{nrqcdl}) as
well as the pNRQCD Lagrangian (\ref{pnrqcdl}) are invariant under a CP plus
$m_1 \leftrightarrow m_2$ transformation.

We will match the Green function $G_{\rm NRQCD}$ defined by 
\bea
G_{\rm NRQCD} 
&=& \langle \vac \vert  \chi^\dagger({\bf x_2},{T/2}) \phi({\bf x}_2,{\bf x}_1;{T/2})\psi({\bf x_1},T/2) 
\nn\\
& & \qquad\qquad\qquad\qquad 
\times \psi^\dagger({\bf y}_1,-T/2)\phi({\bf y}_1,{\bf y}_2;-T/2) 
\chi({\bf y}_2,-T/2) \vert \vac \rangle ,   
\label{nrqcdG}
\eea
with the corresponding Green function in pNRQCD (here and in the rest of
the paper, if not explicitly stated, dependence on ${\bf x}_{1,2}$, ${\bf
p}_{1,2}$ is understood)
\be
\!\!\!\!\!\!\!\!\!\!
G_{\rm pNRQCD} = Z^{1/2} \,\, e^{-iTh_s} \,\, Z^{\dagger 1/2}
\delta^{(3)}({\bf x}_1-{\bf y}_1)\delta^{(3)}({\bf x}_2-{\bf y}_2).
\label{pnrqcdG}
\ee 

\subsection{Calculation in NRQCD}
We expand $G_{\rm NRQCD}$ order by order in $1/m$
$$
G_{\rm NRQCD} = G_{\rm NRQCD}^{(0)} + {1 \over m_1}G_{\rm NRQCD}^{(1,0)} 
+{1 \over m_2}G_{\rm NRQCD}^{(0,1)} + \dots \, . 
$$
Integrating out the fermion fields one gets
\bea
G_{\rm NRQCD}^{(0)} &=& 
 \langle W_\Box \rangle \delta^{(3)}({\bf x}_1-{\bf y}_1)\delta^{(3)}({\bf x}_2-{\bf y}_2),
\label{vsnrqcd}\\
G^{(1,0)}_{\rm NRQCD} &=& {i\over 2} \int_{-T/2}^{T/2}\!\!dt\, 
\langle {\bf D}^2(t) \rangle_\Box \delta^{(3)}({\bf x}_1-{\bf y}_1) \delta^{(3)}({\bf x}_2-{\bf y}_2).
\label{DD} 
\eea
Analogous formulas hold here and in the following for $G_{\rm NRQCD}^{(0,1)}$. 
For simplicity we will not display them. 
The angular bracket $\langle \dots \rangle$ stands for the average value over the 
Yang--Mills action, $W_\Box$ is the rectangular static Wilson loop
$$
W_\Box \equiv {\rm P} \exp\left\{{\displaystyle - i g \oint_{r\times T} \!\!dz^\mu A_{\mu}(z)}\right\},
$$
the angular bracket $\langle \dots \rangle_\Box 
\equiv \langle \dots W_\Box\rangle$ stands for the average over the gauge fields in the presence 
of the static Wilson loop ($\langle 1 \rangle_\Box = \langle W_\Box \rangle)$.  
For further convenience we also define $\langle\!\langle \dots \rangle\!\rangle_\Box 
\equiv \langle \dots W_\Box\rangle / \langle  W_\Box\rangle$.
We use the convention that, if not specified, fields act on the first quark line, 
e.g. ${\bf D}(t) \equiv {\bf D}({\bf x}_1,t)$, 
${\bf E}(t) \equiv {\bf E}({\bf x}_1,t)$, and so on. We have used time reversal:
$\langle {\bf B}(t) \rangle_\Box=-\langle {\bf B}(-t) \rangle_\Box$ to
eliminate the spin-dependent term in Eq. (\ref{DD}). Therefore, we can
already state that no spin-dependent potential appears at $O(1/m)$ \cite{spin1}. 

It is useful to introduce at this point two identities involving covariant derivatives and Schwinger lines:
\begin{eqnarray*}
& &\!\!\!\!\!\!\!
{\rm i)}\, {\bf D}({\bf x},t) \phi(t,{\bf x},t^\prime,{\bf x}) 
= \phi(t,{\bf x},t^\prime,{\bf x})  {\bf D}({\bf x},t^\prime) 
+ ig \int_{t^\prime}^t\!\!dt^{\prime\prime} \, \phi(t,{\bf x},t^{\prime\prime},{\bf x})
{\bf E}({\bf x},t^{\prime\prime})\phi(t^{\prime\prime},{\bf x},t^\prime,{\bf x}) , \\
& &\!\!\!\!\!\!\!
{\rm ii)}\,  {\bf D}({\bf x}_1,t) \phi({\bf x}_1,{\bf x}_2;t) \equiv 
ig {\bf r} \times \int^{1}_{0}ds\,s \, \phi({\bf x}_1,{\bf x}^{\prime}(s);t) 
{\bf B}({\bf x}^{\prime}(s),t)\phi({\bf x}^{\prime}(s),{\bf x}_2;t)  \\
& & \qquad\qquad\qquad\qquad\quad  
+ \phi({\bf x}_1,{\bf x}_2;t) \bfnabla_{{\bf x}_1}, 
\qquad {\bf x}^{\prime}(s)=s {\bf x}_1 + (1-s){\bf x}_2, 
\end{eqnarray*}
where $\phi(t,{\bf x},t^{\prime},{\bf x}) \equiv P \displaystyle\exp
\left\{ -ig\int_{t^\prime}^{t}\!\!dt^{\prime\prime} \, 
A_0({\bf x},t^{\prime\prime})\right\}$. In the first Ref. \cite{spin1}, both
identities i) and ii) were derived. Identity ii) corrects
their Eq. (4.7c). As a  by-product of the above identities, we get  
\begin{equation}
\bfnabla_{{\bf x}_1} \langle W_\Box \rangle = i g \int_{-{T/2}}^{T/2} \!\! dt \, \langle{\bf E }(t) \rangle_\Box 
+  \langle {\bf O}_i(-T/2)\rangle_\Box -  \langle {\bf O}_f(T/2)\rangle_\Box, 
\label{apqua}
\end{equation}
where 
\begin{eqnarray*}
{\bf O}_i(t) &=& ig{\bf r}\times\int^{1}_{0}ds\,s \, \phi({\bf x}_1,{\bf x}^{\prime}(s);t) 
{\bf B}({\bf x}^{\prime}(s),t)\, \phi({\bf x}^{\prime}(s),{\bf x}_2;t), \quad 
{\bf x}^{\prime}(s)= {\bf x}_2 + s{\bf r},\\
{\bf O}_f(t) &=& ig{\bf r}\times\int^{1}_{0}ds\,(1-s) \, 
\phi({\bf x}_2,{\bf x}^{\prime\prime}(s);t) {\bf B} ({\bf x}^{\prime\prime}(s),t) \,
\phi({\bf x}^{\prime\prime}(s),{\bf x}_1;t), \quad 
{\bf x}^{\prime\prime}(s)= {\bf x}_1 - s{\bf r}. 
\end{eqnarray*}
Let us note that time-reversal symmetry gives $\langle ig{\bf E}(t) \rangle_\Box$ $=$  
$\langle ig{\bf E}(-t) \rangle_\Box$ and  
$\langle {\bf O}_i(-T/2) \rangle_\Box$ $=$ $- \langle {\bf O}_f(T/2)\rangle_\Box$. 

Using the above relations, Eq. (\ref{DD}) can be worked out giving 
\begin{eqnarray}
G^{(1,0)}_{\rm NRQCD} 
&=& {i\over 2} \bigg\{{T \over 2} \bfnabla^2_{{\bf x}_1} \langle W_\Box\rangle 
+ {T \over 2} \langle W_\Box\rangle \bfnabla^2_{{\bf x}_1}
+ T\langle {\bf O}_f(T/2)\cdot{\bf O}_i(-T/2)\rangle_\Box
\nonumber \\
& & + i g \!\!\int_{-T/2}^{T/2} \!\!dt \left( {T\over 2} -t\right) 
\langle {\bf O}_f(T/2)\cdot{\bf E}(t)\rangle_\Box 
- i g \!\!\int_{-T/2}^{T/2} \!\!dt \left( {T\over 2} +t\right) 
\langle {\bf E}(t)\cdot{\bf O}_i(-T/2)\rangle_\Box 
\nonumber \\
& & + {g^2\over 2}\int_{-T/2}^{T/2} \!\!dt \int_{-T/2}^{T/2}\!\! dt^\prime 
\vert t -t^\prime \vert \langle {\bf E}(t) \cdot {\bf E}(t^\prime) \rangle_\Box \bigg\}
\delta^{(3)}({\bf x}_1-{\bf y}_1)\delta^{(3)}({\bf x}_2-{\bf y}_2).
\label{g10}
\end{eqnarray}

\subsection{Calculation in pNRQCD}
Let us consider the Green function defined in Eq. (\ref{pnrqcdG}).
Expanding it up to order  $1/m$, we obtain   
$$
G_{\rm pNRQCD} = G_{\rm pNRQCD}^{(0)} + {1\over m_1}G_{\rm pNRQCD}^{(1,0)} 
+{1\over m_2} G_{\rm pNRQCD}^{(0,1)}.
$$
Inserting Eqs. (\ref{hss}) and (\ref{zed}) into Eq. (\ref{pnrqcdG}), we obtain 
\bea 
G_{\rm pNRQCD}^{(0)} &=& 
Z_0 e^{-i V_0 T}\delta^{(3)}({\bf x}_1-{\bf y}_1)\delta^{(3)}({\bf x}_2-{\bf y}_2) 
\label{pnrqcd0}\\
G^{(1,0)}_{\rm pNRQCD} &=&  
Z_0 e^{-iV_0 T} \left\{ {Re[Z_1]\over Z_0} 
-{1\over 2 } \left( \bfnabla\cdot{\bf r}{Z_{1,p}\over Z_0} \right) \right.
\nn \\
& & \!\!\!\!\!\!\!\!\!
- {i\over 2} T (\bfnabla V_0)\cdot{\bf r}{Z_{1,p}\over Z_0} 
+ iT { \bfnabla^2_{{\bf x}_1} \over 2}  - iTV_1 
\nn \\
& & \!\!\!\!\!\!\!\!\!
+ {i T \over 8}\left( 4 {(\bfnabla Z_0)\over Z_0} \cdot \bfnabla_{{\bf x}_1}
+ 2 {(\bfnabla^2 Z_0)\over Z_0} - {(\bfnabla Z_0)^2\over Z_0^2}\right) 
\nn \\
& & \!\!\!\!\!\!\!\!\!
+ {T^2\over 4} \left(2 (\bfnabla V_0)\cdot \bfnabla_{{\bf x}_1} + 
{(\bfnabla Z_0)\over Z_0}\cdot(\bfnabla V_0) + (\bfnabla^2 V_0) \right)
\nn \\
& & \!\!\!\!\!\!\!\!\!
\left. 
-{iT^3 \over 6} (\bfnabla V_0)^2 \right\}
\delta^{(3)}({\bf x}_1-{\bf y}_1)\delta^{(3)}({\bf x}_2-{\bf y}_2), 
\label{pnrqcd}
\end{eqnarray}
where $\bfnabla=\bfnabla_{\bf r}$. 
As in the NRQCD case, we do not display the analogous formulas for
$G_{\rm pNRQCD}^{(0,1)}$. In order to keep Eqs. (\ref{pnrqcd0}) and
(\ref{pnrqcd}) simpler,  we have already used the fact that $Z_0$ and $Z_{1,p}$ can be chosen as real. 
This follows from the matching to $G_{\rm NRQCD}^{(1,0)}$ (\ref{g10}) 
once any constant phase in $Z_0$ is conventionally set to zero.

\subsection{Matching}
At $O(1/m^0)$ we match Eq. (\ref{vsnrqcd}) with Eq. (\ref{pnrqcd0}). We obtain 
\be 
V_0 = \lim_{T\to\infty}{i\over T} \ln \langle W_\Box \rangle.  
\label{v0}
\ee
At $O(1/m)$ we match Eq. (\ref{g10}) with Eq. (\ref{pnrqcd}). We obtain
\begin{eqnarray}
& &V_1 +{1\over 2} (\bfnabla V_0)\cdot{\bf r}{Z_{1,p}\over Z_0} 
= \lim_{T\rightarrow \infty} \Bigg(
- {1 \over 8} \left( { (\bfnabla Z_0) \over Z_0}\right)^2 
+ i {T \over 4}{(\bfnabla Z_0)\over Z_0} \cdot(\bfnabla V_0)  
+ {T^2 \over 12}(\bfnabla V_0)^2 
\nn\\
& &\qquad - {g \over 4}\int_{-T/2}^{T/2} \!\!dt \left\{ \left( 1 -{2t\over T} \right) 
\langle\!\langle {\bf O}_f(T/2)\cdot{\bf E}(t)\rangle\!\rangle_\Box  -\left( 1 +{2t\over T} \right) 
\langle\!\langle {\bf E}(t)\cdot{\bf O}_i(-T/2)\rangle\!\rangle_\Box \right\}
\nn\\
& &\qquad - {1 \over 2} \langle\!\langle{\bf O}_f(T/2){\bf O}_i(-T/2)\rangle\!\rangle_\Box
- {g^2\over 4 T}\int_{-T/2}^{T/2} \!\! dt \int_{-T/2}^{T/2} \!\!dt^\prime \vert t -t^\prime \vert 
\langle\!\langle {\bf E}(t) \cdot {\bf E}(t^\prime)\rangle\!\rangle_\Box \Bigg). 
\label{result}
\end{eqnarray}
A similar expression with neither end-point string contributions nor
normalization factors has been obtained in \cite{thesis}.
The right-hand side of Eqs. (\ref{v0}) and (\ref{result}) can be shown to be  
finite in the $T\to\infty$ limit. This is not obvious from the point of view of a pure Wilson-loop calculation 
in the spirit of Ref. \cite{spin1}, as is apparent from the difficulties met by
the author of \cite{thesis} (the normalization factors are crucial in order to get a finite expression). 
For this reason, and because such kind of arguments may become relevant to future analyses 
of Wilson-loop correlators, we mention here the relevant steps of the proof. \\
{\it a)} Inserting the identity operator  $\sum |n\rangle^{(0)}
{}^{\,\,(0)}\langle n|$ into the Wilson-loop average, the latter may be
written as  
$$
\langle W_\Box \rangle = \sum_n e^{-i E^{(0)}_n T} a_n^2.
$$
Doing the same for the average $\langle {\bf E}(t)\rangle_\Box$ one obtains   
$$
\langle {\bf E}(t)\rangle_\Box = \sum_{n} e^{-iE^{(0)}_n T} a_n^2 
{}^{\,\,(0)}\langle n\vert {\bf E} \vert n\rangle^{(0)} 
+ \sum_{n \neq m} e^{-i(E^{(0)}_n+E^{(0)}_m) {T\over 2} + i(E^{(0)}_n-E^{(0)}_m) t} 
a_n a_m {}^{\,\,(0)}\langle n\vert {\bf E} \vert m\rangle^{(0)}. 
$$
{\it b)} End-point strings containing the operators ${\bf O}_f$ and ${\bf O}_i$  
select intermediate states with quantum numbers different from the singlet.  
This can be checked directly  on  the definitions of these operators.  
As an immediate consequence, correlators containing the operators ${\bf O}_f$ and ${\bf O}_i$ 
in Eq. (\ref{result}) vanish in the $T\to\infty$ limit and do not contribute to the potential. \\
{\it c)} From Eq. (\ref{apqua}) and time-inversion invariance of the
chromoelectric field,  it follows that 
\begin{eqnarray*}
- (\bfnabla V_0) &  = &  {}^{\,\,(0)}\langle 0 \vert g {\bf E} \vert 0\rangle^{(0)}, \\
(\bfnabla Z_0) &=& 2 \sum_{n\neq 0} a_0 a_n {{}^{\,\,(0)}\langle 0 \vert g {\bf E} \vert n\rangle^{(0)} 
\over E_n -E_0}.
\end{eqnarray*} 
{\it d)} Inserting the identity operator into the correlator 
$\langle {\bf E}(t) \cdot {\bf E}(t^\prime)\rangle_\Box$, it may be written as
$$
\langle {\bf E}(t) \cdot {\bf E}(t^\prime) \rangle_\Box 
=\sum_{n,m,s} a_n a_m  {}^{\,\,(0)}\langle n | {\bf E} |s \rangle^{(0)} 
{}^{\,\,(0)}\langle s | {\bf E} |m \rangle^{(0)}
e^{-i(E^{(0)}_n+E^{(0)}_m){T\over 2} }e^{i (E^{(0)}_n-E^{(0)}_s)t} e^{i(E^{(0)}_s-E^{(0)}_m)t^\prime},
$$
for $t>t^\prime$.
With the above points {\it a)-d)} it is easy to show that the right-hand side 
of Eqs. (\ref{v0}) and (\ref{result}) is finite in the large-$T$ limit. Their explicit expressions read 
\be
V_0=E_0^{(0)}
\ee
\bea
&&V_1 +{1\over 2} (\bfnabla V_0)\cdot{\bf r}{Z_{1,p}\over Z_0} 
= \lim_{T\rightarrow \infty} \Bigg( - {1 \over 8} \left( { (\bfnabla Z_0) \over Z_0}\right)^2 
+ i {T \over 4}{(\bfnabla Z_0)\over Z_0} \cdot(\bfnabla V_0) + {T^2 \over 12}(\bfnabla V_0)^2 
\nn\\
& &\qquad 
- {g^2\over 4 T}\int_{-T/2}^{T/2} \!\! dt \int_{-T/2}^{T/2} \!\!dt^\prime \vert t -t^\prime \vert 
\langle\!\langle {\bf E}(t) \cdot {\bf E}(t^\prime)\rangle\!\rangle_\Box \Bigg)
\nn\\
&&\qquad= {1\over 2} \sum_{n\neq 0} \left\vert 
{ {}^{\,\,(0)} \langle n | g{\bf E} | 0 \rangle^{(0)} \over E_0^{(0)} - E_n^{(0)}} \right\vert^2
+ (\bfnabla E_0^{(0)})\sum_{n\neq 0} {a_n \over a_0}
{ {}^{\,\,(0)} \langle n | g{\bf E} | 0 \rangle^{(0)} \over (E_0^{(0)} - E_n^{(0)})^2}\, .
\label{v1z1p}
\eea
Analogously, we can obtain matching expressions for the normalization
factors. At $O(1/m^0)$, the result is $Z_0=|a_0|^2$. At $O(1/m)$ a more
complicated equality is obtained, which involves a combination of $Z_1$ and
$Z_{1,p}$. The fact that we have two equations and three independent
functions: $Z_1$, $Z_{1,p}$ and $V_1$ at $O(1/m)$ makes the
determination of $V_1$ ambiguous. This ambiguity is intrinsic in the sense
that any value of $Z_{1,p}$ and $V_1$ that fulfils Eq. (\ref{v1z1p}) will lead
to the same physics (as far as one consistently works at higher orders in
$1/m$), and has to do with the fact that a quantum-mechanical Hamiltonian is defined 
up to time-independent unitary transformations (see Appendix). 
As a consequence, the ambiguity in the definition
of $V_1$ also gives 
rise to ambiguities in the definition of the potentials of order $O(1/m^2)$ and higher. 
In the next section we will fix $Z_{1,p}$ 
(and then $V_1$) by imposing an extra matching condition, which   
will prove to be particularly convenient. 

\section{The quantum-mechanical matching}
\label{qmmatch}
In the previous section we have done the matching comparing Green
functions in NRQCD and in pNRQCD. In this section the comparison is
between states and matrix elements in NRQCD and in pNRQCD. The
calculation, in terms of states, will be closer in philosophy to the usual quantum
mechanics calculations in perturbation theory \cite{Sakurai} (see also
\cite{swa}). Moreover, the whole
procedure will share some similarities to the adiabatic approximation and
the Born--Oppenheimer method as used in atomic physics calculations \cite{Messiah}.
The underlying assumption is that the difference of energies among states labelled with different
$n$ is much larger than the difference of energies among states
labelled with the same $n$. In our case, since we only aim at correctly
reproducing the ground state spectrum, 
we only need that the splitting between the ground state and the first gluonic
excitation be larger 
than the typical splitting of the states of the ground state (taken of $O(mv^2)$ by
definition). This is nothing but the condition we have assumed throughout the
present paper.

$H$ is not diagonal in the basis of $H^{(0)}$ ($|\underbar{n}\rangle^{(0)}$) 
with respect to the $n$ labelling. We consider instead a basis of states, labelled as 
\be
|\underbar{n}; {\bf x}_1 ,{\bf x}_2\rangle ,
\label{basis1}
\ee 
such that the Hamiltonian $H$ is diagonal with respect to these states 
\be
\langle \underbar{m}; {\bf x}_1 ,{\bf x}_2|H| \underbar{n}; {\bf y}_1
,{\bf y}_2\rangle =\delta_{nm}E_{n}({\bf x}_1 ,{\bf x}_2,{\bf p}_1,{\bf p}_2)
\delta^{(3)} ({\bf x}_1 -{\bf y}_1)\delta^{(3)} ({\bf x}_2 -{\bf y}_2) \,,
\label{cond1}
\ee
where $E_{n}({\bf x}_1 ,{\bf x}_2,{\bf p}_1,{\bf p}_2)$ is an analytic function 
in ${\bf p}_1$, ${\bf p}_2$, and such that they are normalized as 
\be
\langle \underbar{m}; {\bf x}_1 ,{\bf x}_2| \underbar{n}; {\bf y}_1,{\bf y}_2\rangle =
\delta_{nm}\delta^{(3)} ({\bf x}_1 -{\bf y}_1)\delta^{(3)} ({\bf x}_2 -{\bf y}_2) \,.
\label{cond2}
\ee
Conditions (\ref{cond1}) and (\ref{cond2}) give 
\be
H| \underbar{n}; {\bf y}_1 ,{\bf y}_2\rangle
= \int d^3{\bf x}_1d^3{\bf x}_2| \underbar{n}; {\bf x}_1 ,{\bf x}_2\rangle 
E_{n}({\bf x}_1 ,{\bf x}_2,{\bf p}_1,{\bf p}_2)
\delta^{(3)} ({\bf x}_1 -{\bf y}_1)\delta^{(3)} ({\bf x}_2 -{\bf y}_2) \,.
\label{cond1p}
\ee
A set of states $|\underbar{n}\rangle$ and an operator $E_{n}$ that satisfy 
Eqs. (\ref{cond2}) and (\ref{cond1p})
can be obtained from the static solutions $|n\rangle^{(0)}$ and $E_{n}^{(0)}$ to any desired
order of accuracy by working out formulas analogous to the ones used in standard quantum mechanics 
perturbation theory \cite{Sakurai}. If we write $|\underbar{n}\rangle$ and $E_{n}$ as an expansion in $1/m$,
\be
|\underbar{n} \rangle = |\underbar{n} \rangle^{(0)} +{1 \over m_1}|\underbar{n} \rangle^{(1,0)}
+{1 \over m_2}|\underbar{n}\rangle^{(0,1)} + \dots, 
\label{fullstate}
\ee
\be
E_n =E_n^{(0)}+ {1 \over m_1}E_n^{(1,0)}+{1 \over m_2}E_n^{(0,1)} + \dots,
\label{fullenergy}
\ee
we obtain at $O(1/m)$ for $n=0$ (when not specified, states and energies are calculated 
in ${\bf x}_1$ and ${\bf x}_2$): 
\bea
|\underbar{0} \rangle^{(1,0)} &=& {1\over E_0^{(0)} -H^{(0)}} \sum_{n\neq 0}
\int d^3{\bf x}^{\prime}_1\,d^3{\bf x}^{\prime}_2\,
|\underbar{n}; {\bf x}^\prime_1 ,{\bf x}^\prime_2 \rangle^{(0)} 
{}^{\,\,(0)}\langle \underbar{n}; {\bf x}^\prime_1 ,{\bf x}^\prime_2 |
H^{(1,0)} |\underbar{0} \rangle^{(0)}
\nn \\
&=& -{1\over E_0^{(0)} -H^{(0)}}
\sum_{n\neq 0} \left\{ { {}^{\,\,(0)} \langle n | g{\bf E} | 0 \rangle^{(0)} \over
E_0^{(0)} - E_n^{(0)}} \cdot \bfnabla_{{\bf x}_1} + {1\over 2} \left(\bfnabla_{{\bf x}_1} 
{ {}^{\,\,(0)} \langle n | g{\bf E} | 0 \rangle^{(0)} \over
E_0^{(0)} - E_n^{(0)}} \right) \right. \nn\\
& & \qquad\qquad \left. - {1\over 2} \sum_{j\neq 0,n} 
{ {}^{\,\,(0)} \langle n | g{\bf E} | j \rangle^{(0)} 
{}^{\,\,(0)} \langle j | g{\bf E} | 0 \rangle^{(0)} 
\over (E_j^{(0)} - E_n^{(0)}) (E_j^{(0)} - E_0^{(0)})}\right\}|\underbar{n} \rangle^{(0)}, 
\label{QM2} 
\eea
\bea
& &E_0^{(1,0)} \delta^{(3)} ({\bf x}_1 -{\bf y}_1)\delta^{(3)} ({\bf x}_2 -{\bf y}_2) =
{}^{\,\,(0)} \langle \underbar{0}|H^{(1,0)} |\underbar{0} \rangle^{(0)} 
\nn \\
& & \qquad\qquad = \left( -{{\bfnabla_{{\bf x}_1}}^2 \over 2} + 
{1\over 2} \sum_{n\neq 0} \left\vert { {}^{\,\,(0)} \langle n | g{\bf E} 
| 0 \rangle^{(0)} \over E_0^{(0)} - E_n^{(0)}} \right\vert^2 \right)
\delta^{(3)} ({\bf x}_1 -{\bf y}_1)\delta^{(3)} ({\bf x}_2 -{\bf y}_2). 
\label{QM1}
\eea
These equations may be derived from the identities
$$
\hspace{-4.2cm}\hbox{a)} \quad 
{}^{\,\,(0)} \langle n | {\bf D}_{{\bf x}_1} | n \rangle^{(0)} =
\bfnabla_{{\bf x}_1},  
$$
which follows from symmetry considerations and 
$$
\hbox{b)} \quad 
{}^{\,\,(0)} \langle n | {\bf D}_{{\bf x}_1} | j \rangle^{(0)} = 
{ {}^{\,\,(0)} \langle n | g{\bf E}({\bf x}_1) | j \rangle^{(0)} \over
E_n^{(0)} - E_j^{(0)}} \quad \forall \, n\neq j, 
$$
which follows from explicit calculation. Analogous formulas hold for the
antiparticle contribution.

\subsection{Matching}
The aim of pNRQCD is to describe the behaviour of $|\underbar{0}\rangle$. The
integration of high excitations is trivial using the basis (\ref{basis1}) since,
in this case, they are decoupled from $|\underbar{0}\rangle$. Therefore, the
matching of NRQCD to pNRQCD is basically to rename things in a way such that pNRQCD 
reproduces the matrix elements of NRQCD for the ground state. The matching conditions in this
formalism read as follows:
\be
|\underbar{0}\rangle = S^{\dagger} |\vac \rangle \quad {\rm and} \quad E_0({\bf x}_1,{\bf x}_2, 
{\bf p}_1, {\bf p}_2) = h_s({\bf x}_1,{\bf x}_2, {\bf p}_1, {\bf p}_2)  \,.
\label{singlet}
\ee
Using Eq. (\ref{singlet}) and Eq. (\ref{Sdef}) we obtain the
normalization factor, given the interpolating field, 
\begin{equation}
Z^{1/2}({\bf x}_1,{\bf x}_2, {\bf p}_1,{\bf p}_2)
\delta^{(3)} ({\bf x}_1 -{\bf y}_1)\delta^{(3)} ({\bf x}_2 -{\bf y}_2)
= \langle \vac| \chi^\dagger({\bf x}_2) \phi({\bf x}_2,{\bf x}_1) 
\psi({\bf x}_1)| \underbar{0}; {\bf y}_1,{\bf y}_2 \rangle .  
\label{Z}
\end{equation}
It is explicit now that $h_s$ and, hence, the potential, fixed by Eq. 
(\ref{singlet}), depend neither on the normalization factors nor on the specific
shape of the end-point strings. Moreover,
Eqs. (\ref{singlet}) and (\ref{Z}) provide matching conditions at any finite order in $1/m$. 

From Eqs. (\ref{singlet}) and (\ref{Z}) one could draw the conclusion
that the ambiguity in the computation of $V_1$, discussed in the previous 
section, has disappeared. This is not the case. Actually, Eqs. (\ref{fullstate}) and (\ref{fullenergy}) 
with Eqs. (\ref{QM2}) and (\ref{QM1}) only give one of the possible solutions
of Eqs. (\ref{cond1p}) and (\ref{cond2}). 
Indeed, these equations do not completely fix the state $|\underbar{n}\rangle$.
In standard quantum mechanics the state is fixed up to an arbitrary constant phase. 
In our case, since we only diagonalize in the ${n}$ space, this phase becomes
a unitary operator\footnote{Note that the phase is not completely arbitrary if we demand the state
$|\underbar{n}\rangle$ to coincide with the unperturbed state
$|\underbar{n}\rangle^{(0)}$ in the limit $1/m \rightarrow 0$. This constrains
$O_n(x,p)$ to smoothly go to zero in the limit $1/m \rightarrow 0$.}; 
given a state $|\underbar{n}\rangle$ that fulfils
Eqs. (\ref{cond1p}) and (\ref{cond2}), this means that the states 
$$
\int d^3{\bf x}^{\prime}_1 \, d^3 {\bf x}^{\prime}_2\, 
|\underbar{n},{\bf x}^{\prime}_1,{\bf x}^{\prime}_2\rangle e^{iO_n({\bf x}^{\prime}_1,
{\bf x}^{\prime}_2, {\bf p}^{\prime}_1,{\bf p}^{\prime}_2)}
\delta^{(3)} ({\bf x}^{\prime}_1 -{\bf x}_1)\delta^{(3)} ({\bf x}^{\prime}_2 -{\bf x}_2), 
$$ 
with $O_n^{\dagger}=O_n$, still fulfil Eq. (\ref{cond2}) and the
Hamiltonian is still diagonal in $n$ with 
$$
E_n \rightarrow e^{iO_n}\,E_n\,e^{-iO_n} \,.
$$
This freedom reflects the fact that $h_s$ 
is defined by the matching up to an arbitrary unitary transformation 
or, which is the same, up to an arbitrary unitary field redefinition:
$$
h_s \rightarrow e^{iO_0}\,h_s\,e^{-iO_0},
$$
and so does the $Z$. In order to make calculations easier, we have taken advantage of this
freedom by fixing the relative phase between $|\underbar{0}\rangle$ 
and $|\underbar{0}\rangle^{(0)}$ following the standard 
choice of quantum mechanics perturbation theory. With this choice, we 
obtain Eqs. (\ref{QM2}) and (\ref{QM1}); as we will see below, this will allow
us to obtain a compact expression for $V_1$ in terms of Wilson loops. Furthermore,
the procedure can be easily generalized at any finite order in
$1/m$. 

From Eqs. (\ref{QM2}) and (\ref{Z}) we can perform the matching at $O(1/m)$. We obtain 
\begin{eqnarray} 
& &\!\!\!Z_0^{1/2} = a_0, \label{z0}\\
& &\!\!\!{Z_1\over Z_0^{1/2}} = \sum_{n\neq 0} {a_n \over E_0^{(0)} - E_n^{(0)}} \left\{ 
 \left( \bfnabla_{{\bf x}_1}  
{ {}^{\,\,(0)} \langle n | g{\bf E} | 0 \rangle^{(0)} \over E_0^{(0)} - E_n^{(0)}} \right)
- 2\left( { {}^{\,\,(0)} \langle n | g{\bf E} | 0 \rangle^{(0)} 
\over (E_0^{(0)} - E_n^{(0)})^2} \right) (\bfnabla_{{\bf x}_1}E_0^{(0)}) \right. \nn\\
& & \qquad\qquad 
\left. 
+ \sum_{j\neq 0,n} { {}^{\,\,(0)} \langle n | g{\bf E}| j \rangle^{(0)} 
{}^{\,\,(0)} \langle j | g{\bf E}| 0 \rangle^{(0)} 
\over (E_j^{(0)} - E_n^{(0)}) (E_j^{(0)} - E_0^{(0)})} \right\},
\label{z1}\\
& &\!\!\!{Z_{1,p}{\bf r}\over Z_0^{1/2}} = 2 \sum_{n\neq 0} a_n 
{ {}^{\,\,(0)} \langle n | g{\bf E}| 0 \rangle^{(0)} \over
(E_0^{(0)} - E_n^{(0)})^2}.
\label{z1p}
\end{eqnarray}
From Eqs. (\ref{QM1}) and (\ref{singlet}) we get, up to $O(1/m)$:
\bea
V_0(r)&=&E_0^{(0)}(r),
\label{v0s}\\
V_1(r) &=&  {1\over 2} \sum_{n\neq 0} \left\vert{ {}^{\,\,(0)} \langle n | 
g{\bf E} | 0 \rangle^{(0)} \over E_0^{(0)} - E_n^{(0)}} \right\vert^2.
\label{v1s}
\eea

At first sight, this result seems of limited practical utility, since it depends on the exact 
and complete solution of the bound state at order $1/m^0$. Fortunately this is not the case. 
In fact, Eq. (\ref{result}) suggests a very simple form for the $1/m$ potential,
\be
V_1 = \lim_{T\to\infty}\left(
- {g^2\over 4 T}\int_{-T/2}^{T/2} \!\! dt \int_{-T/2}^{T/2} \!\!dt^\prime
\vert t -t^\prime \vert \bigg[ \langle\!\langle {\bf E}(t) \cdot {\bf E}(t^\prime)\rangle\!\rangle_\Box 
- \langle\!\langle {\bf E}(t)\rangle\!\rangle_\Box \cdot
\langle\!\langle {\bf E}(t^\prime)\rangle\!\rangle_\Box  \bigg] \right).
\label{v1E}
\ee
Using the same technique as outlined in {\it a)-d)} of Sec. 3.3, it can be proved that 
Eq. (\ref{v1E}) is finite and equal  to Eq. (\ref{v1s}). 

Equation (\ref{v1E}) is the main result of the present work. It gives
the leading $1/m$ correction to the heavy-quark 
potential expressed in terms of chromoelectric field insertions in a
static Wilson loop, thus avoiding the explicit computation of the
normalization factors. The potential appears, therefore, in the same form  as
the potential calculated in Refs. \cite{spin1,spin2,cor} and is suitable for lattice evaluations
similar to those performed in \cite{latpot}. Moreover, the above expression
expanded in $\als$ gives the full perturbative series of the $1/m$ potential in the regime where
this expansion makes sense (i.e. $m v \gg \lQ$). In the next section we will
calculate from it the leading non-vanishing perturbative contribution to the
$1/m$ potential, which will coincide with the results of Refs. \cite{MY,H}. For the calculation of 
some higher order perturbative corrections to the $V_1$ potential  we refer to \cite{nnnll}.
Finally, it is interesting to note that 
Eq. (\ref{v1E}) also holds if the correlators are evaluated on a Wilson
loop with infinitely large time extension (i.e. if we make the limit $T \rightarrow \infty$ in the
Wilson loops for $t$ and $t^{\prime}$ finite, and then evaluate the integrals). 
This observation greatly simplifies the perturbative calculation and may
also do so for non-perturbative ones.

\section{Applications}
\label{application}
In this section we will consider Eq. (\ref{v1E}) in perturbative QCD and in quenched QED. 
In the first case we will evaluate the contribution to the potential 
at the leading non-vanishing order in $\als$.
In the second we will show that the potential $V_1$ vanishes exactly. 
This may be relevant to several Gaussian models used in the phenomenology of the non-perturbative 
dynamics of the strong interaction. We will shortly comment on this.

\begin{figure}[htb]
\vskip 0.5truecm
\makebox[0truecm]{\phantom b}
\put(40,0){\epsfxsize=12.0truecm\epsffile{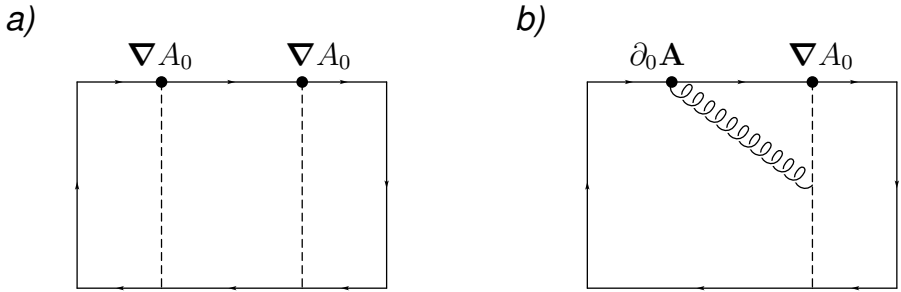}}
\put(90,85){$\bfnabla A_0$}
\put(150,85){$\bfnabla A_0$}
\put(280,85){$\partial_0 {\bf A}$}
\put(340,85){$\bfnabla A_0$}
\vskip 0.2truecm
\caption{{{\it Dashed lines indicate Coulombic exchanges, curly lines transverse gluons. Graph 
{\tt a)} is the disconnected graph cancelling the $T^2 (\bfnabla V_0)^2/12$ term in Eq. (\ref{result}). 
Graph {\tt b)} (and the symmetric graph which is understood) gives the only non-vanishing contribution to the $1/m$
potential at order $\als^2$ in the 
Coulomb gauge. In other gauges additional diagrams may be present.}}}
\label{perV1}
\end{figure}

\subsection{Perturbative QCD}
We perform the calculation in the Coulomb gauge. Since Eq. (\ref{v1E}) is a
gauge-independent quantity, the result will hold in any gauge. 
At order $\als$, the right-hand side of Eq. (\ref{v1E}) only gives a self-energy type of
contribution, since the chromoelectric fields are inserted on the 
same quark line. The first non-vanishing contribution to the potential (i.e. depending on ${\bf r}$) 
appears at order $\als^2$. Three types of diagrams arise:\\
{\it i) Disconnected graphs.} An example is shown by graph {\tt a)} of Fig. \ref{perV1}. 
These graphs cancel in the difference of the right-hand side of Eq. (\ref{v1E}). \\
{\it ii) Connected graphs with end-point strings contributions.} Non-disconnected graphs involving gluons attached 
to the end-point strings vanish in Eq. (\ref{v1E}). As noticed in the previous section the $ T
\rightarrow \infty$ limit 
can also be performed in the correlators before doing the time integrals in
Eq. (\ref{v1E}). This turns out to be quite useful here, 
making the irrelevance  of these graphs to the potential manifest.\\
{\it iii) Connected graphs with no end-point strings contributions.} The only
non-vanishing graph contributing to Eq. (\ref{v1E}) is 
graph {\tt b)} of Fig. \ref{perV1}. The analogous graph involving the triple-gluon 
vertex, but with two transverse gluons attached to the chromoelectric fields
on the quark line, can be shown to vanish in the limit of Eq. (\ref{v1E}) by explicit calculation.\\
Therefore, considering all relevant contributions (which reduce to graph {\tt b)} of Fig. \ref{perV1})
we get from Eq. (\ref{v1E}) at order  $\als^2$ 
\be
V_1^{\rm pert} =  - C_FC_A{\als^2 \over 4\,r^2}.  
\label{V1pertres}
\ee
Equation (\ref{V1pertres}) coincides with the result of Refs. \cite{MY,H} (which may also 
be obtained  using the rules of \cite{pos})
and with the non-Abelian part of the perturbative calculation 
of the $1/m$ potential done in \cite{pert}. In Ref. \cite{pert} also an
Abelian contribution to the $1/m$ potential 
is found. We have stressed throughout the paper that there is no unique way to define the 
$1/m$ potential  and, therefore, different matching procedures are, in general, expected to 
give different expressions for it. One may suspect that  this is due to the fact  that in 
Refs. \cite{MY,H} the Coulomb gauge is used whereas in \cite{pert} the Feynman gauge is used.
However, this  has more to do with the freedom we have in quantum mechanics to change the form 
of the Hamiltonian by a unitary transformation without changing the physics, rather than with the 
gauge fixing dependence itself.  What typically happens in perturbative calculations, which match 
non gauge invariant Green functions, is that depending on the gauge one uses, one gets a different 
form of the potential. The different forms are equivalent and can be obtained  from each other by 
unitary transformations.  
More precisely, from the discussion in the
Appendix it follows that $V_1^{\rm pert}/m$, at $O(\als^2)$, can be 
rewritten in terms of an $O(\als/m^2)$ potential\footnote{For the general,
  unequal mass, case the Abelian term is of the type $1/(m_1+m_2)$ and the field redefinition discussed in the Appendix would transform it into a $1/(m_1m_2)$ term.}. However, whereas the Abelian
piece found in \cite{pert} can be obtained from the general expressions for the
$O(1/m^2)$ potential \cite{cor}, this is not so for the non-Abelian piece in Eq. (\ref{V1pertres}), 
since there is not non-Abelian contribution at tree
level. Therefore, we conclude that Eq. (\ref{v1E}) represents a genuine
new potential not considered in the past in the non-perturbative
parametrizations of the QCD potential in terms of Wilson loops.  
Finally, let us mention that, to our knowledge, $V_1^{\rm pert}$  
has never been computed using Wilson loops before.

\subsection{Quenched QED}
Let us consider quenched QED. In this situation the Wilson-loop average is exactly known 
(it reduces to Gaussian integrals, see for instance \cite{kogut}) and can be written as 
\be 
\langle W_\Box \rangle = 
\exp\left\{-\displaystyle{g^2\over 2} \int_0^1 ds  \int_0^1 ds^\prime
\int_{-T/2}^{T/2} \!\!dt \int_{-T/2}^{T/2} \!\!dt^\prime r_i r_j f_{ij}(t-t^\prime, (s-s^\prime){\bf r})\right\},
\label{gau}
\ee
where $f_{ij}(t,{\bf r}) = f_{ji}(t,{\bf r})$ $=$ $ \langle E_i(t,{\bf r})E_j(0,{\bf 0})\rangle$.
As a direct consequence (for a derivation, see for instance \cite{fs}) we obtain
\be
\langle\!\langle {\bf E}(t) \cdot {\bf E}(t^\prime)\rangle\!\rangle_\Box
- \langle\!\langle {\bf E}(t)\rangle\! \rangle_\Box
\cdot \langle\!\langle {\bf E}(t^\prime) \rangle\!\rangle_\Box =
f_{ii}(t-t^\prime,{\bf 0}).
\label{gau1}
\ee
Therefore, since the term $f_{ii}(t-t^\prime,{\bf 0})$ is a self-energy-type 
contribution, the potential contribution of $V_1$ (as defined in Eq. (\ref{v1E}))
vanishes exactly in
quenched QED. The same result is obtained in Ref. \cite{pos}.

Several QCD vacuum models \cite{rev,cor1,mod} seem to approximate the Wilson-loop 
long-range non-perturbative dynamics with expressions analogous to Eq. (\ref{gau}). 
The above considerations made for quenched QED may be relevant to them. 
As a matter of fact, $1/m$ corrections do not seem  to show up there. Therefore, it is tempting 
to generalize the exact result of quenched QED to Gaussian models of 
the QCD vacuum. Nevertheless some words of caution are needed. 
A model consists in an approximation of QCD, which is supposed to coincide with 
a relevant limit on the QCD dynamics. This limit is unknown by definition.  
Therefore, how to implement it is, in the practice of several models,  
not well established. While it is clear that starting from a Gaussian expression 
for the Wilson loop a relation like Eq. (\ref{gau1}) is going to hold, 
it is not guaranteed that $1/m$ potentials will not show up if the Gaussian 
approximation of the Wilson loop is implemented at some intermediate step of the 
calculation\footnote{The reader may, for instance, compare the formulas 
of Refs. \cite{spin1,spin2} with the predictions of some Gaussian models 
discussed in Ref. \cite{mod}.}.

\section{Conclusions}
We have obtained an exact and non-perturbative expression for $V_1$, the $Q\bar{Q}$ 
potential at order $1/m$. This expression is shown in Eq. (\ref{v1E}). The
perturbative calculation of it up to order $\als^2$ 
agrees with \cite{MY,H} and
is consistent with the one-loop contribution of \cite{pert}. However, the existence of a non-perturbative 
contribution at order $1/m$ was, to our knowledge,  not considered before
in the literature, be it in phenomenological applications \cite{rev} 
or in attempts at obtaining the non-perturbative potentials from QCD \cite{spin1,spin2,cor}.
We note that via a unitary transformation the $1/m$ terms can, 
in certain circumstances,
be reshuffled in $1/m^2$ (and higher) terms in the potential (see Appendix). 
When these circumstances apply (i.e. when $V_1 \ll m^2 v^2$), our results
 do not have immediate 
consequences on phenomenological
 models where the full set of $1/m^2$ potentials are considered and input as ans\"atze functions. 
However, when they do not (i.e. when $V_1 \sim m^2 v^2$), our results imply that a $1/m$ potential 
should be included in those models. In either case,
they are extremely important for the attempts at obtaining the potentials 
from lattice QCD, since, upon doing the above-mentioned reshuffling, the $1/m$ potential 
given by Eq. (\ref{v1E}) generates $1/m^2$ terms different from the ones calculated so far \cite{cor}. 

The most promising way to have an estimate of the non-perturbative behaviour 
of the correlators appearing in Eq. (\ref{v1E}) is by a lattice simulation. 
In practice this can already be done from the lattice data of Ref. \cite{latpot}, since 
the correlators we get are of the same type. Such lattice data could be of interest 
for at least two reasons. If the potential found on the lattice happens 
to be small in the long range, it would support the Abelian dominance picture. In the short range 
it should show the interplay of the perturbative (known) region with the non-perturbative one.

When there are no additional US degrees of freedom, the adopted non-perturbative procedure can 
be easily generalized to the evaluation of higher-order terms in the inverse mass expansion \cite{PV}. 
The inclusion of US degrees of freedom in the non-perturbative regime will be discussed elsewhere.

\vspace{0.3cm}

\noindent 
{\bf Acknowledgements.} 
N.B. acknowledges the TMR contract No. ERBFMBICT961714, A.P. the TMR contract No. ERBFMBICT983405, 
J.S. the contracts AEN98-031 (Spain) and 1998SGR 00026 (Catalonia).
N.B., J.S. and A.V. acknowledge the program ``Acciones Integradas 1999-2000'', project No. 13/99.
N.B. and A.V. thank the University of Barcelona for hospitality while part of this work 
was carried out. N.B. and A.V. acknowledge Dieter Gromes for 
interesting discussions and for calling their attention to Ref. \cite{thesis}.

\vfill\eject

\appendix
\Appendix{}
In this appendix we show that there exists a unitary transformation that reshuffles 
mo\-men\-tum-in\-dependent $1/m$ terms in $1/m^2$ momentum-dependent and momentum-independent
terms. Let us consider the Hamiltonian 
\begin{equation}
H = {p^2\over 2 m} + V_0(r) + {V_1(r)\over m}. 
\label{hap1}
\end{equation}
The unitary transformation 
$$
U = \exp\left( - {i\over m} \{ {\bf W}(r), {\bf p} \}\right), 
$$
transforms $H \to H^\prime = U^\dagger \, H \, U$.
More explicitly, under the condition
$\{ {\bf W}, {\bf p}\} \ll m $ (which is necessary in order to maintain 
the standard form of the leading terms in the Hamiltonian, i.e. a kinetic term plus a 
velocity independent potential)
 $H^\prime$ reads 
\begin{eqnarray}
H^\prime &=& {p^2\over 2 m} + V_0 + {V_1\over m} + {2\over m}{\bf W}\cdot ({\bfnabla} V_0) 
+ {2\over m^2}{\bf W}\cdot ({\bfnabla} V_1) 
\nn\\
& & + {2\over m^2}W^i(\bfnabla^i W^j (\bfnabla^j V_0)) - {1\over 2 m^2}\{p^i,\{ p^j, (\nabla^i W^j) \}\} 
+ O\left({1\over m^3}\right).
\label{hap2}
\end{eqnarray}
By choosing 
$$
{\bf W} = -{1\over 2} V_1 {{\bfnabla} V_0 \over ({\bfnabla} V_0)^2}
$$
all the $1/m$ potentials in the Hamiltonian (\ref{hap2}) disappear. 
The price to pay is the appearance of new terms at order $1/m^2$ (and higher). 
Of course, the leading size  of these new terms is the same as 
 the original $V_1/m$. In particular, 
since ${\bf p}$, $\bfnabla$, $\sim m v$ and $V_0 \sim mv^2$,  
we get ${\bf W} \sim V_1/(m^2 v^3)$ and $\{p^i,\{p^j,(\nabla^i W^j)\}\}/m^2 \sim V_1/m$. 
The size of the remaining $1/m^2$ induced terms is $V_1^2/(m^3 v^2)$
and hence, it depends on the size of $V_1/m$, which is a priori unknown.
On general grounds the maximum size of $V_1$ is given by the largest available scale,
 namely $mv$, and hence at most $V_1\sim m^2 v^2$. 
Reasoning in the same way, the $1/m^2$ potentials (calculated via the quantum-mechanical matching in 
\cite{PV}) are not bigger than $mv^3$. However, from the condition
$\{ {\bf W}, {\bf p}\} \ll m $, it follows 
that the reshuffling of the $1/m$ potential to $1/m^2$  potentials may be done in the way above only 
if $V_1\ll m^2 v^2$. As a consequence, all the terms of $O(V_1^2/(m^3v^2))$
have a size much 
smaller than $m v^2$. More specifically, if, for instance, $V_1 \sim m^2 v^3$, then the terms of 
$O(V_1^2/(m^3v^2))$ are of order $mv^4$ and hence suppressed by a factor $v$ with respect to 
the original $1/m$ potential (as well as with respect to the $1/m^2$ potentials obtained from 
the quantum-mechanical matching). We note that in perturbation theory 
($v\sim \als$) $V_1 \sim m^2 v^4$ and, therefore the terms of  $O(V_1^2/(m^3v^2))$ 
are suppressed by a factor $v^2$.

\vfill\eject

\end{document}